\documentclass[fleqn,usenatbib]{mnras}

\usepackage{newtxtext,newtxmath,graphicx}
\usepackage{subfigure}

\usepackage[T1]{fontenc}

\DeclareRobustCommand{\VAN}[3]{#2}
\let\VANthebibliography\thebibliography
\def\thebibliography{\DeclareRobustCommand{\VAN}[3]{##3}\VANthebibliography}

\usepackage{graphicx}	
\usepackage{amsmath}	
\usepackage{amssymb}	
\usepackage{algorithm}
\usepackage{algorithmicx,algpseudocode}
\usepackage{diagbox}

\title[CS-WFS]{Compressive Shack-Hartmann Wavefront Sensor based on Deep Neural Networks}

\author[Peng et al.]{
Peng Jia,$^{1, 2, 3}$ \thanks{E-mail: robinmartin20@gmail.com (JP)}
Mingyang Ma,$^{1}$ \thanks{E-mail: mamingyang260@163.com (MM)}
Dongmei Cai,$^{1}$  \thanks{E-mail: dm\underline{ }cai@163.com (DM)}
Weihua Wang,$^{1}$
Juanjuan Li$^{1}$
and Can Li$^{4}$
\\
$^{1}$College of Physics and Optoelectronics, Taiyuan University of Technology, Taiyuan, 030024, China\\
$^{2}$Key Laboratory of Advanced Transducers and Intelligent Control Systems, Ministry of Education and Shanxi Province, \\
Taiyuan University of Technology, Taiyuan, 030024, China\\
$^{3}$Department of Physics, Durham University, South Road, Durham DH1 3LE, UK\\
$^{4}$College of Information and Computer Science, Taiyuan University of Technology, Taiyuan, 030024, China
}

\date{Accepted XXX. Received YYY; in original form ZZZ}

\pubyear{2015}

\begin{document}
\label{firstpage}
\pagerange{\pageref{firstpage}--\pageref{lastpage}}
\maketitle

\begin{abstract}
The Shack-Hartmann wavefront sensor is widely used to measure aberrations induced by atmospheric turbulence in adaptive optics systems. However if there exists strong atmospheric turbulence or the brightness of guide stars is low, the accuracy of wavefront measurements will be affected. In this paper, we propose a compressive Shack-Hartmann wavefront sensing method. Instead of reconstructing wavefronts with slope measurements of all sub-apertures, our method reconstructs wavefronts with slope measurements of sub-apertures which have spot images with high signal to noise ratio. Besides, we further propose to use a deep neural network to accelerate wavefront reconstruction speed. During the training stage of the deep neural network, we propose to add a drop-out layer to simulate the compressive sensing process, which could increase development speed of our method. After training, the compressive Shack-Hartmann wavefront sensing method can reconstruct wavefronts in high spatial resolution with slope measurements from only a small amount of sub-apertures. We integrate the straightforward compressive Shack-Hartmann wavefront sensing method with image deconvolution algorithm to develop a high-order image restoration method. We use images restored by the high-order image restoration method to test the performance of our the compressive Shack-Hartmann wavefront sensing method. The results show that our method can improve the accuracy of wavefront measurements and is suitable for real-time applications.
\end{abstract}

\begin{keywords}
techniques: image processing -- instrumentation: adaptive optics -- techniques: high angular resolution
\end{keywords}


\section{Introduction}
\label{sec:Intro} 
The Shack-Hartmann wavefront sensor (SH-WFS) is a wavefront sensing device that normally consists of a lenslet array and an opto-electronic sensor. In real applications, the SH-WFS firstly divides the wavefront into several sub-apertures and light in each sub-apertures will be focused by a lenslet. Wavefront slope measurements in each sub-aperture will then be evaluated according to displacements of spots formed by each lenslet. Finally, wavefront difference maps could be obtained through integration of wavefront slope measurements \citep{Platt2001}.\\

The principle of the SH-WFS indicates us that as the number of sub-apertures increases, the SH-WFS can measure wavefront difference maps with higher spatial resolution. Meanwhile, because the atmospheric turbulence is highly variable in the temporal domain, the SH-WFS needs to work in very high frame-rate. With more sub-apertures and higher frame-rates, there would be less photons that could be received by each sub-aperture in each frame. Considering the brightness of guide stars in adaptive optics systems is low, there is obvious a trade-off between the highest speed and highest spatial resolution that a SH-WFS can measure. In real applications, if the atmospheric turbulence is strong or the reference target is dim, slope measurements calculated from spot images with low signal to noise ratio would lead to unreliable wavefront measurements \citep{thomas2006comparison}. To solve this problem, many new methods are proposed \citep{gilles2008constrained, schreiber2009laser, basden2015sensitivity, townson2015improved, anugu2018peak, adam2019improving}.\\

Because wavefront difference maps are normally continuous surfaces, they can be reconstructed with undersampled signals represented by different basis functions. This concept, known as compressive sensing \citep{donoho2006compressed}, has already been applied in holography \citep{clemente2013compressive} and single pixel camera \citep{duarte2008single}. In these applications, original signals can be reconstructed with only a small number of undersampled measurements.\\

For SH-WFS, \citet{polans2014compressed} propose to represent wavefront difference maps with Zernike polynomials \citep{hosseini2009derivative} and have shown that the compressive sensing could increase the speed of wavefront measurements. Because the atmospheric turbulence is random media that satisfies a particular statistical property \citep{roggemann1996}, new sparse basis functions, such as the golden section sparse basis function \citep{juanjuan2018sparse}, are further proposed to reduce reconstruction error. Compressive SH-WFS methods mentioned above could reconstruct wavefront difference maps with only tens of percentage of all sub-apertures. Because slope measurements introduced by spot images with low signal to noise ratio (SNR) is the main error of SH-WFS, the compressive wavefront sensing method indicates us a possibility that we could choose spot images with high SNR to better reconstruct wavefront difference maps. With only a small amount of slope measurements, it would be possible to increase the speed and the spatial resolution for a SH-WFS at the same time.\\

The basis functions and the corresponding reconstruction algorithms are two important factors for a compressive SH-WFS. Normally it would cost scientists a long time to design adequate basis functions and corresponding reconstruction algorithms. In essence, design of basis functions and reconstruction algorithms is the procedure of finding a complex function that could map undersampled slope measurements to original wavefront difference maps. In recent years, deep neural networks (DNN) have become a successful problem solver in representation of complex functions \citep{liu2017survey}. For wavefront sensing problems, \citet{nishizaki2019deep, andersen2019neural} develop DNNs to learn mapping functions between wavefront represented by Zernike coefficients and images blurred by the wavefront. \citet{dubose2020intensity} propose to train a DNN that can map slope measurements and intensity measurements to original wavefront difference maps, which can increase the detection ability of SH-WFS for strong atmospheric turbulence. In this paper, we take advantage of the complex function representation ability of DNNs and propose to design a compressive SH-WFS method.\\

The compressive Shack-Hartmann wavefront sensing method (CS-WFS) proposed in this paper uses a DNN to learn mapping functions between sparsely sampled slope measurements and original wavefront difference maps. In real applications, spot images with high signal to noise ratio will be selected to calculate slope measurements and then the trained DNN could reconstruct original wavefront difference maps from these measurements. To accelerate algorithm development, we introduce a drop-out layer to simulate the compressive sensing process during the training stage. This paper is organized as follows. In Section \ref{sec:Method}, we will introduce the CS-WFS. In Section \ref{sec:decon}, we will integrate the CS-WFS with the deconvolution from wavefront sensing algorithm \citep{primot1990deconvolution} to develop high-order image deconvolution method. We use deconvolved images to test the effectiveness of our method. In Section \ref{sec:con}, we will make our conclusions and anticipate our future works.\\

\section{Compressive Shack-Hartmann Wavefront Sensing Method Based on deep neural networks}
\label{sec:Method} 
The classical SH-WFS approximates continuous wavefronts with discrete slope measurements from equally distributed sub-apertures and reconstruct wavefronts from slope measurements. The classical wavefront reconstruction algorithm is the Southwell algorithm. However when the wavefront error is large or the number of photons obtained by each sub-aperture is small, SNR of spot images would be too low to obtain reliable slope measurements \citep{adam2019improving}. In this case, reconstructing wavefront difference maps with slope measurements calculated from spot images of high SNR would be a possible way to assure accuracy of SH-WFS. Because atmospheric turbulence induced aberrations are stochastic, sub-apertures that have spot-images with high SNR would distribute randomly and unequally. Therefore we need a compressive SH-WFS method to reconstruct wavefront difference maps.\\

The concept of the compressive SH-WFS was firstly proposed by \citet{polans2014compressed} with the name of SPARZER. The SPARZER is capable to reconstruct original wavefront difference maps represented by Zernike polynomials with only $5\%$ of randomly sub-sampled SH-WFS measurements. However, since wavefront difference maps are represented by Zernike polynomials in the SPARZER, the spatial resolution of wavefront measurement is limited by the maximal orders of Zernike polynomials. Besides, the SPARZER requires around 30 iterations to reconstruct the wavefront, which limits its application in astronomy observations that require real-time and high-order wavefront measurements.\\

Meanwhile, DNNs can be trained straightforwardly in an end-to-end way and could be directly used for wavefront sensing without numerical iterations. In recent years, DNNs have been used in slope estimation, wavefront reconstruction and wavefront estimation from images \citep{mello2014artificial, li2018centroid, hu2019learning, xu2020wavefront, hu2020deep}. In this paper, we will build a straightforward compressive SH-WFS method (CS-WFS) based on DNN compressive sensing concepts \citep{lu2018convcsnet, wu2019deep}. We will discuss details of the CS-WFS in Section \ref{sec:CSWF}.\\

\subsection{The Structure of the CS-WFS Method}
\label{sec:CSWF}
\begin{figure*}
	\includegraphics[width=1.5 \columnwidth]{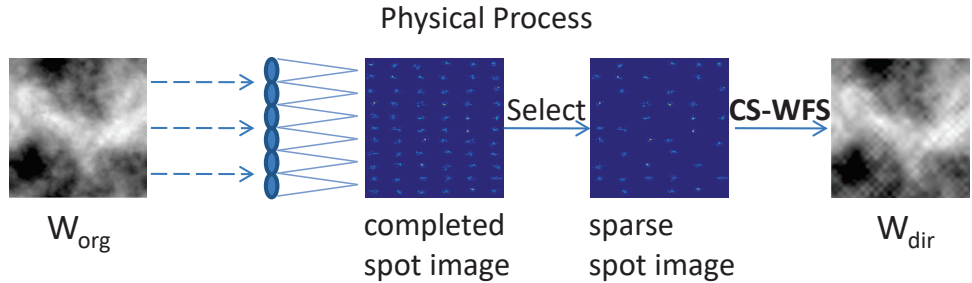}
    \caption{The physical process of the CS-WFS method. The first step is the same as that of SH-WFS, which will sample the wavefront $W_{org}$ with lenslets to form complete spot images. Then the CS-WFS will select spot images with enough SNR as sparse spot images to obtain effective slope measurements. At last these effective images will be sent to the deep neural network in CS-WFS to reconstruct wavefront maps $W_{dir}$.}
    \label{fig:1}
\end{figure*}

The physical process of the CS-WFS in real applications is shown in figure \ref{fig:1}. Comparing with classical SH-WFS, the CS-WFS has an additional SNR based sub-aperture selection step. This step is used to guarantee that slope measurements are reliable for further wavefront reconstruction. Because there are many different slope calculation methods \citep{nicolle2004improvement, gilles2008constrained}, the SNR should be defined accordingly. In this paper, we select the classical centre-of-gravity algorithm and use SNR defined in equation \ref{eq:1} as selection criterion. \\

\begin{equation}
    SNR = \frac{signal-noise}{\sigma_{noise}}
	\label{eq:1}
\end{equation}
The $SNR$ is calculated by each sub-apertures. $signal$ is the summation of grey scale values of pixels which are larger than a predefined threshold (one standard deviation larger than the mean grey scale value for this paper). $noise$ represents the average grey scale of the background and $\sigma_{noise}$ is the standard deviation of the background.\\

Steps of the CS-WFS are shown in algorithm \ref{algorithm1}. It includes four main steps:\\
1. Calculation of SNR from spot images. In this step, we will calculate SNR of spot images for each sub-apertures.\\
2. Selection of sub-apertures. After calculation of SNR of spot images of all sub-apertures, we could either select slope measurements that are larger than a predefined threshold $SNR_t$ (for relatively weak atmospheric turbulence) or slope measurements from spot images with $N_s$ largest SNR (for strong atmospheric turbulence) for wavefront reconstruction.\\ 
3. Calculation of slope measurements. Spot images from selected sub-apertures will be used to calculate the centre of spot images. Then we can obtain slope measurements from the centre of spot images.\\
4. Reconstruction of wavefront difference maps. With DNNs in CS-WFS, we can reconstruct wavefront difference maps from these slope measurements.\\
\begin{algorithm}
\caption{Main steps of the CS-WFS method}
\label{algorithm1}
\begin{algorithmic}[1]
\Require 
\Statex $SNR_t$\ Threshold of SNR of spot images
\Statex $N_{s}$\ Minimal number of sub-apertures
\Ensure
\Statex $F_{DNN}(SLOPE)$\ DNN that can map sparse slope measurements $SLOPE$ to wavefront difference maps
\Statex
\State Evaluating SNR of spot images in each sub-apertures as $L_{SNR}$
\State Sorting $L_{SNR}$ and calculating number of sub-apertures $N_{SNR}$ that have spot images with SNR larger than $SNR_t$
\Statex \If{$N_{SNR}\le N_{s}$}
\State Selecting sub-apertures with the biggest $N_{s}$ SNR to calculate slope measurements ($SLOPE$) from spot images
\Statex \Else
\State Selecting sub-apertures with the biggest $N_{SNR}$ SNR to calculate slope measurements ($SLOPE$) from spot images
\EndIf
\State Reconstructing wavefront path difference maps from $SLOPE$ with $F_{DNN}(SLOPE)$
\end{algorithmic}
\end{algorithm}

\begin{figure*}
	\includegraphics[width=1.5 \columnwidth]{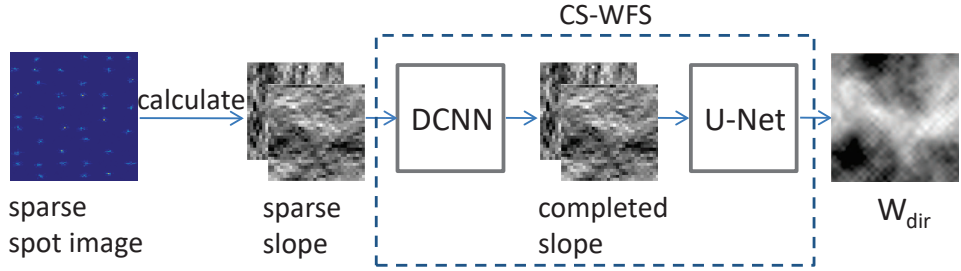}
    \caption{The flow chart of the CS-WFS. The sparse spot images will be used to calculate sparse slope measurements along x and y directions. Then these sparse slope measurements will be used to calculate complete slope measurements with the DCNN and then the complete slope measurements will be used to reconstruct wavefronts by the U-Net.}
    \label{fig:2}
\end{figure*}

The flowchart of algorithms of the CS-WFS method is shown in figure \ref{fig:2}. The main part of the CS-WFS is the DNN that could reconstruct wavefront difference maps from sparsely sampled slope measurements. To make results obtained by DNNs comprehensible, we use two sequentially connected DNNs to reconstruct wavefront difference maps. The first DNN is a deep convolutional neural network (DCNN), which is mainly used to estimate complete slope measurements from sparsely sampled slope measurements. The second DNN is a U-Net, which is mainly used to reconstruct wavefront difference maps from complete slope measurements.\\

\begin{figure*}
	\includegraphics[width=1.5 \columnwidth]{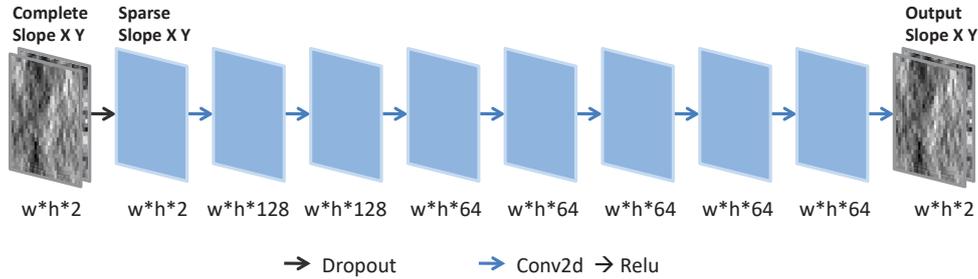}
    \caption{The structure of the DCNN that are used to estimate complete slope measurements from sparse slope measurements. It includes 8 convolutional layers and a drop-out layer in front of them is used to simulate the compressive sensing process during training.}
    \label{fig:3}
\end{figure*}

\begin{figure*}
	\includegraphics[width=1.8 \columnwidth]{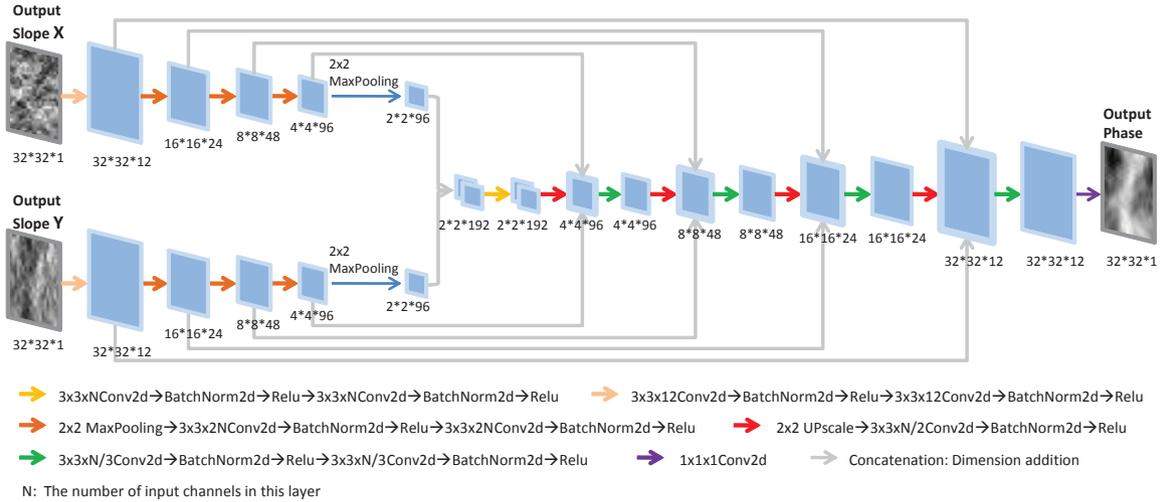}
    \caption{The structure of the U-Net for the wavefront reconstruction. Complete slope measurements of x and y directions will be transferred into two separate branches and then outputs of these two branches will be merged together to estimate wavefront difference maps.}
    \label{fig:4}
\end{figure*}

The structure of the DCNN in the CS-WFS is shown in figure \ref{fig:3}. The DCNN includes 8 convolutional layers to estimate complete slope measurements from sparsely sampled slope measurements. In front of the first layer is a drop-out layer \citep{hinton2012improving}. The drop-out layer will randomly set a percentage of weights as zero to prevent over-fitting of a neural network. In this paper we use it to simulate the compressive sensing process. During the training stage, we set the drop-out rate to a predefined value. When the complete slope measurements are input into the DCNN network, the predefined proportional (drop-out rate) slope measurements value are set to zero through the drop-out layer to obtain the sparse slope measurements. Then the DCNN will be able to learn the function that could estimate complete slope measurements from sparsely sampled slope measurements.\\
The structure of the U-Net is shown in figure \ref{fig:4}. It is used to reconstruct wavefront difference maps from complete slope measurements. The U-Net is modified from the structure proposed in \citet{dubose2020intensity}. In this paper, the U-Net merges complete slope measurements along x and y directions to reconstruct wavefront difference maps.\\

\subsection{Training of the CS-WFS Method}
\label{sec:trainCSWF} 
We use a high-fidelity atmospheric turbulence phase screen model \citep{jia2015simulation,jia2015real} and an adaptive optics system simulation platform \citep{Basden2018} to generate atmospheric turbulence phase screens and corresponding spot images to train and test the CS-WFS. The Fried parameter ($r_0$) is set to 0.06, 0.10, 0.14, 0.18 meter and the diameter of the telescope is set to 1 metre. In this paper, the SH-WFS has $32\times 32$ sub-apertures and we assume there are no photo-electronics noise in the SH-WFS. With this configuration, we could obtain noise free spot images. Besides, we resize atmospheric turbulence phase screens to $32\times 32$ pixels as wavefront difference maps that should be obtained by our CS-WFS method.\\

Complete slope measurements are calculated by noise free spot images through the centre-of-gravity algorithm. The U-Net is trained by complete slope measurements and their corresponding resized phase screens. Meanwhile, we use complete slope measurements as the training set of the DCNN. During the training stage, we set the drop-out rate as 0.9 in the drop-out layer of the DCNN, which means only $10\%$ slope measurements can transmit to following layers (compressive rate 0.9). The drop-out rate refers to percentage of signals that are directly set to zero and the compressive rate refers to percentage of slope measurements that are directly set as zero. We find that if the DCNN is trained with drop-out rate of 0.9  (compressive rate 0.9), the CS-WFS can reconstruct all wavefronts that have smaller or larger compressive rates. Therefore we set the compressive rate as 0.9 during training stage. After training, compressive slope measurements will be sent into the DCNN and the drop-out rate will be set to zero (all slope measurements will be transmitted to next layers - compressive rate equals to 0.0). Then the DCNN could reconstruct wavefronts from these compressive slope measurements.\\

The DCNN and the U-Net can be linked sequentially and trained together or separately. We find it is more efficient to train the DCNN and the U-Net together. During training stage, the initial learning rate of the DCNN is 0.001 and that of the U-Net is 0.0001. The batch size is 100 and we use 130 epochs to train them. The wavefront slope and wavefront phase for training are processed by linear normalization. The loss function of the CS-WFS is defined in equation \ref{eq:2}, where $err_C$  is the loss function of the DCNN and $err_G$ is the loss  function of the U-Net. For both of these two DNNs, we use the mean square error (MSE) as the loss function and Adam algorithm to optimize these two DNNs \citep{kingma2014adam}.After training, the CS-WFS can restore wavefronts from slope measurements with 21 ms in a computer with Nvidia Tesla K40c GPU under double float numerical accuracy. The speed can be further increased with DNNs implemented in FPGAs with numerical accuracy of int8 or int16.\\
\begin{equation}
    err_{Plus}  = err_{C}/200+err_{G}.
	\label{eq:2}
\end{equation}

\subsection{Performance of the CS-WFS Method in Measurement of Wavefronts}
\label{sec:CSWFResult} 
In this section, we generate 200 atmospheric turbulence phase screens for different blur levels ($r_0$ of 6 cm, 12 cm and 18 cm) to test the performance of the CS-WFS. The input of the network for testing is the wavefront slope that has not been normalized. We use the root mean square error ($RMSE$) of residual wavefronts defined below as the evaluation function.\\
\begin{equation}
    RMSE = \sqrt {\frac{1}{N}\sum_{i=1}^{n}(W_{org}-W_{dir})^2}.
	\label{eq:3}
\end{equation}
where $W_{org}$ and $W_{dir}$ are original and reconstructed wavefronts, N is the total number of pixels that are used to represent the wavefront. The RMSE is defined with unit of radians. In this paper, we assume the reconstructed wavefront has the same dimension as that of the slope measurements. Therefore $N$ equals to $32\times 32$ (1024) pixels.\\

\begin{figure*}
	\includegraphics[width=1.2 \columnwidth]{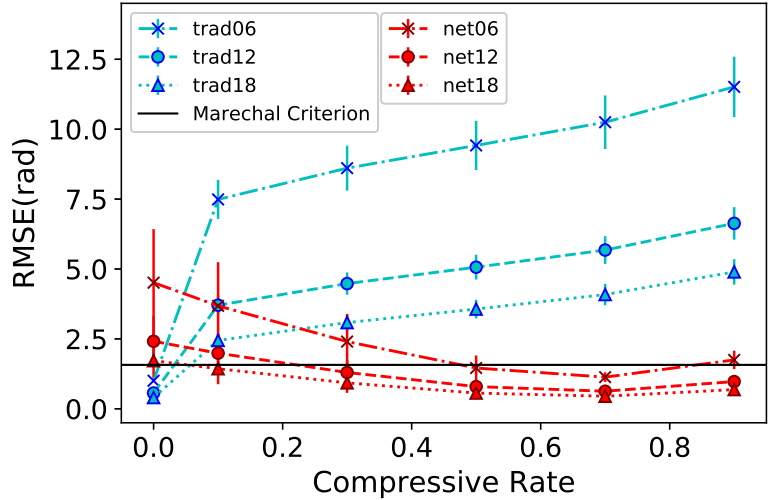}
    \caption{Comparison of the RMSE value of residual wavefront obtained by the classical method and the compressive sensing method. net stands for the compressive wavefront reconstruction method. trad stands for the classical compressive wavefront reconstruction method. The number at the end of net or trad stands for the Fried parameter in centimetre. Compressive rate stands for percentage of slope measurements that are removed from all measurements.}
    \label{fig:5}
\end{figure*}

We firstly test the robustness of the CS-WFS under different compressive rates. We train a CS-WFS with a compressive rate of 0.9 (compressive rate of 0.9: $10\%$ sub-apertures are used in applications) and use slope measurements with compressive rate from 0.0 to 0.9 to test performance of the CS-WFS.  The results are shown in figure \ref{fig:5}. We can find that the CS-WFS is a robust wavefront measure method: although we use a predefined compressive rate to train the CS-WFS, the trained CS-WFS can reconstruct wavefront with different compressive rates. In this figure, we can also find that the classical method has good performance when all sub-apertures are used for reconstruction. When the number of effective sub-apertures reduces, the performance of the classical wavefront reconstruction algorithm will drop down.\\

Then we test the CS-WFS algorithms with noisy wavefront slope measurements. To better control different noise levels in measurements from SH-WFS, we generate noisy slope measurements through adding random Poisson noise in noise free slope measurements. We will adjust magnitude of noise and compare noisy wavefront slopes with original slopes. When the difference between noisy slopes and original slopes is larger than $10\%$ of PV values of original wavefront slopes, we will set that as "noisy wavefront slopes" or "error slope" or "incorrect slope". The noise level is set as fraction of error slope measurements of all slope measurements in one frame.\\

\begin{figure*}
\centering
\subfigure[diameter = 1m , $r_0 = 6 cm$]{
\begin{minipage}[t]{1\columnwidth}
\centering
\includegraphics[width=1\columnwidth]{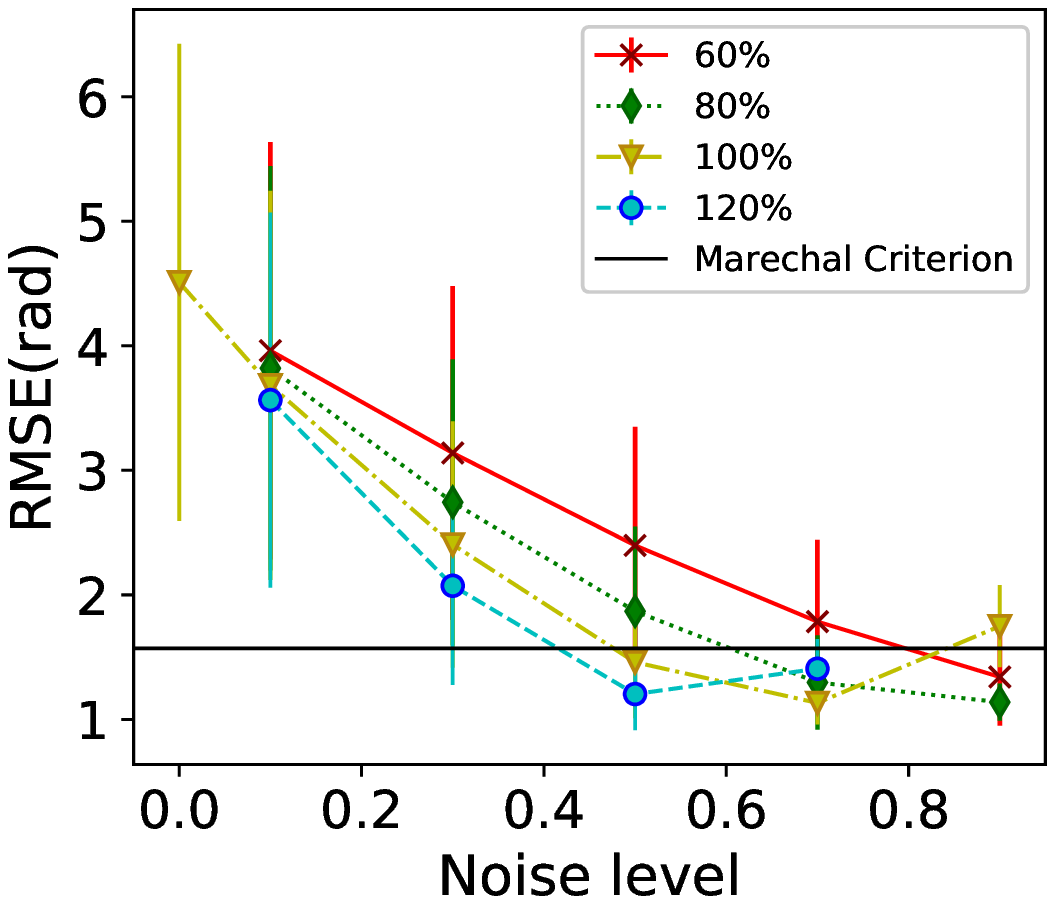}
\end{minipage}%
}%
\subfigure[diameter = 1m , $r_0 = 12 cm$]{
\begin{minipage}[t]{1\columnwidth}
\centering
\includegraphics[width=1\columnwidth]{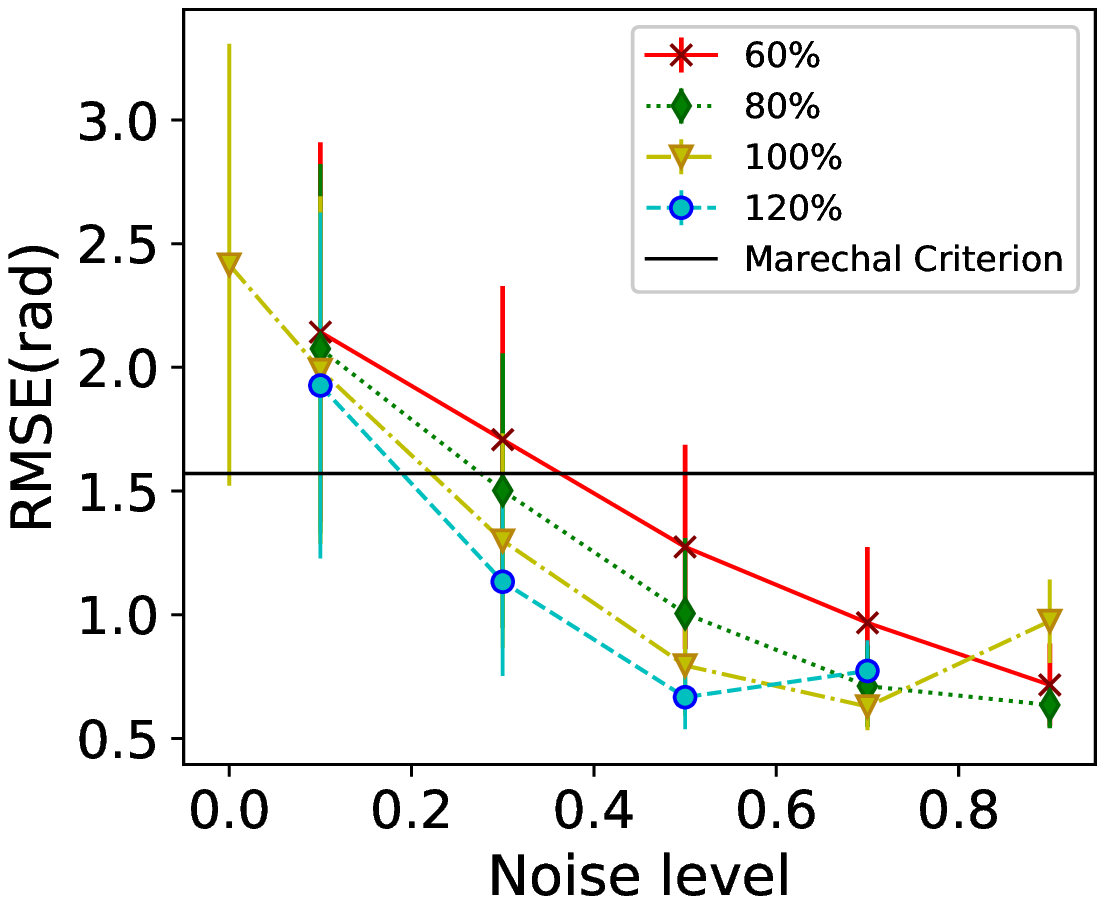}
\end{minipage}%
}\\
\subfigure[diameter = 1m , $r_0 = 18 cm$]{
\begin{minipage}[t]{1\columnwidth}
\centering
\includegraphics[width=1\columnwidth]{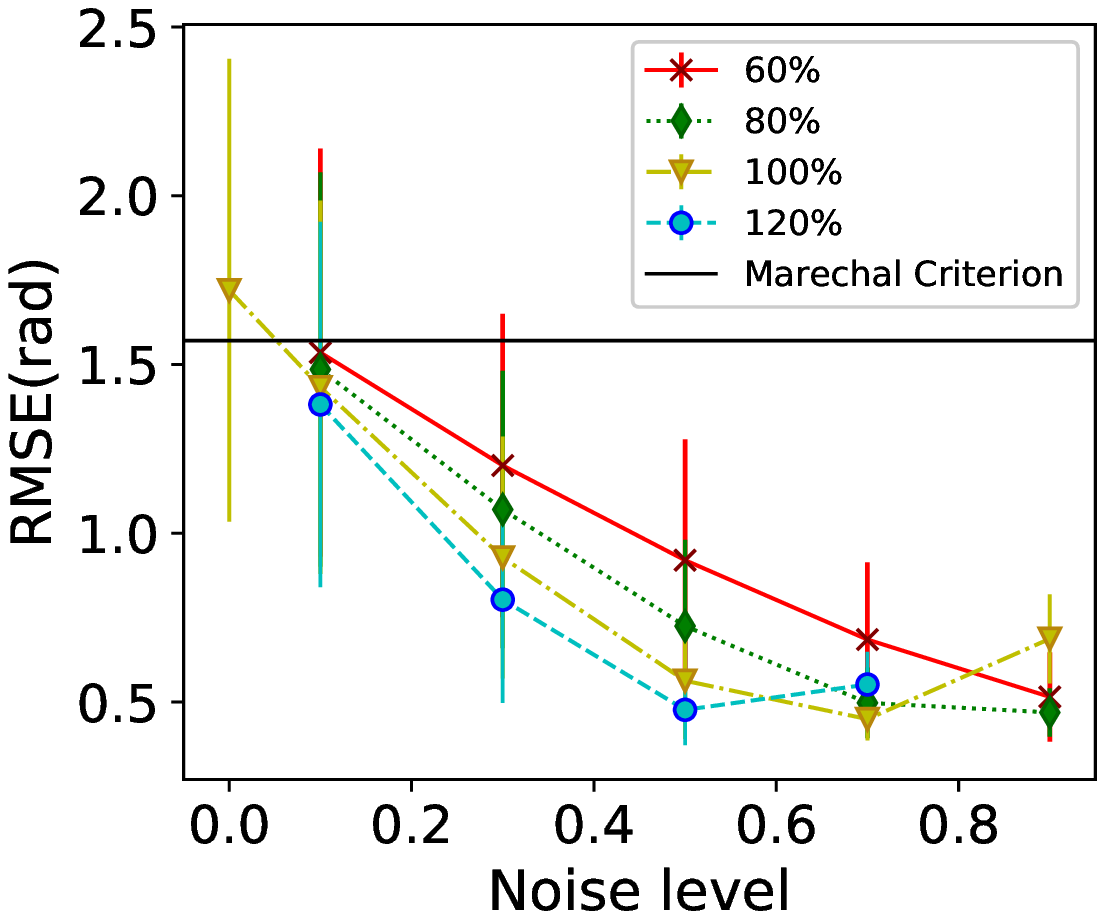}
\end{minipage}
}%
\centering
\caption{RMSE of Residual wavefront reconstructed from different number of selected sub-apertures for different noise level (percentage of sub-apertures with incorrect slope measurements). (a) stands for Fried parameter of 6 cm. (b) stands for Fried parameter of 12 cm and (c) stands for Fried parameter of 18 cm. The part of RMSE less than Marechal Criterion is considered to have reached the diffraction limit.}
\label{fig:6}
\end{figure*}

Due to the lack of original slope measurements in real applications, the method to determine whether the slope measurements are error measurements is hard to use. Therefore error slopes will be selected according to the SNR of spot images. If the SNR of spot images is used as selection criterion,  three different situations would occur: all correct slope measurements are selected, fewer correct slope measurements are selected and some incorrect slope measurements are selected. We test the performance of the CS-WFS method under these three different situations. As shown in figure \ref{fig:6}, there are different percentage of sub-apertures which have incorrect slope measurements. We set the drop-out rate as 0.9 during the training stage. In real applications, we choose percentage of sub-apertures that are removed from observation data from $60 \%$ to $120 \%$ of the number of sub-apertures that have incorrect slope measurements. We can find that the CS-WFS is robust and can reconstruct wavefronts with more or less effective slope measurements. Besides, there is a weak relation that as the number of sub-apertures with incorrect slope measurements increases, it would be better to remove more sub-apertures.\\

\section{Application of the CS-WFS in Deconvolution from Wavefront Sensing}
\label{sec:decon} 
Deconvolution from wavefront sensing (DWFS) is a powerful low-cost high-resolution imaging technique \citep{primot1990deconvolution, welsh1995signal, mugnier2001myopic, jefferies2011deconvolution}. During observations, the DWFS firstly records wavefront measurements and short-exposure images simultaneously. Then the DWFS deconvolves short-exposure images with the point spread function (PSF) estimated from wavefront measurements. Because the DWFS uses light from the observed target for wavefront sensing and imaging, its observation ability for dim targets are limited. In this part, we take advantage of the higher spatial resolution and better reconstruction accuracy of the CS-WFS method and integrate the DWFS and the CS-WFS as a new image restoration method.\\

\begin{figure*}
	\includegraphics[width= 1.5 \columnwidth]{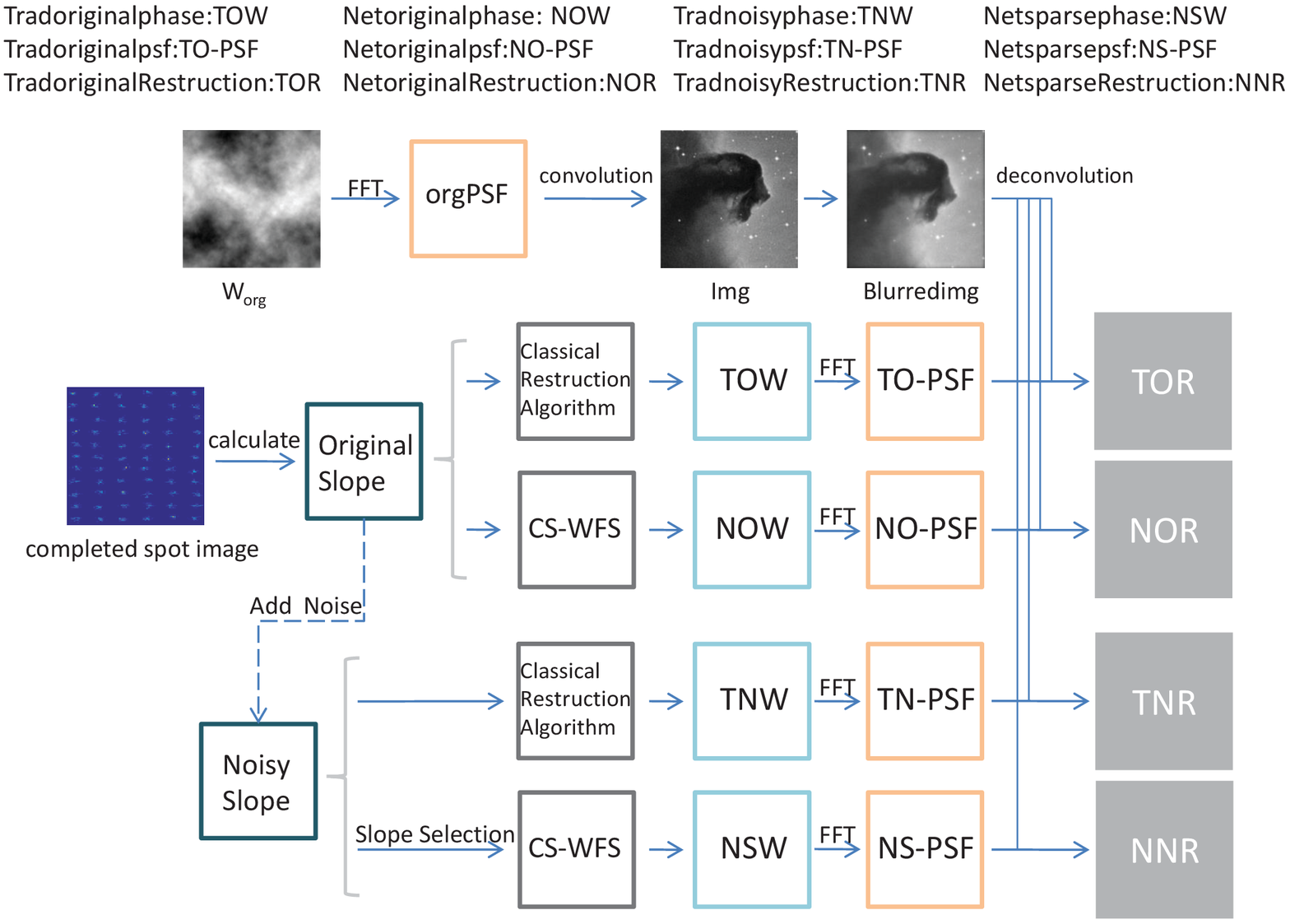}
    \caption{The flowchart of the CS-WFS and the DWFS in this paper. Original wavefronts are used to generate blurred images. Four PSFs are estimated from original wavefronts or wavefront measurements. They are PSF estimated by the classical wavefront reconstruction algorithm from noise free complete slope measurements (TO-PSF), PSF estimated by the CS-WFS algorithm from noise free complete slope measurements (NO-PSF), PSF from noisy complete slope measurements (TN-PSF) and PSF from noisy compressive slope measurements (NS-PSF).}
    \label{fig:7}
\end{figure*}

The flowchart of the new image restoration method is shown in figure \ref{fig:7}. Blurred images used in this section are generated through convolution of short exposure PSFs and high resolution images. The short exposure PSFs are also generated by Monte Carlo simulation platform \citep{Basden2018} with a high fidelity atmospheric turbulence phase screen model \citep{jia2015simulation, jia2015real}. We set the Fried parameter ($r_0$) as 6 cm, 12 cm and 18 cm. Meanwhile, we set the diameter of telescope as 1 m. Original wavefronts are stored as $W_{org}$. We set the number of sub-apertures in SH-WFS as $32 \times 32$. Through Monte Carlo simulation, we could obtain spot images from wavefronts. With classical wavefront reconstruction algorithm, we obtain wavefront optical difference maps from these spot images as $W_{dir}$. Then we can obtain the PSF (TO-PSF) from $W_{dir}$ through far field diffraction. Meanwhile, we can also obtain the PSF (NO-PSF) from complete slope measurements with the CS-WFS method. PSFs mentioned above are obtained by noise free complete slope measurements. TN-PSFs are obtained by classical wavefront reconstruction method from complete noisy slope measurements. NS-PSFs are obtained by CS-WFS method from compressive noisy slope measurements.\\

According to noise levels, we select slope measurements with different compressive rates and obtain NS-PSFs and NO-PSFs. All these PSFs are used as prior condition for deconvolution algorithm. In this paper, we use the maximum likelihood deconvolution algorithm from MATLAB to deconvolve blurred images and use estimated PSFs as prior PSF. We calculate SSIM (Structural Similarity defined in MATLAB) between deconvolved images and original images to evaluate the quality of deconvolved images. The results are shown in table \ref{tab:imgSSIM1}. We can find that the CS-WFS is robust to different compressive rates (NNR is larger than TNR).\\

\begin{table*}
	\centering
	\caption{SSIM of restored images with different percentages of slope measurements that are incorrect. Fried parameter stands for Fried parameter in centimetre. Compressive rate stands for fraction of slope measurements that are removed for wavefront reconstruction. Blurred Image stands for SSIM of blurred images. Restored Image stands for SSIM of restored images with corresponding blurred PSFs as prior information. TOR and NOR stand for SSIM of images restored with PSF obtained from the classical method and the compressive wavefront reconstruction method with wavefront measurements of no noise. TNR and NNR stand for SSIM of images restored with the classical and the compressive wavefront reconstruction algorithms with noisy wavefront measurements. We find that in the case of no noise, the classical algorithm has better restoration performance than the network method. Meanwhile, when the noise level increases, the network method has better restoration performance than the classical algorithm. And as the noise level increases, the network method is more robust than the classical algorithm.}
	\label{tab:imgSSIM1}
	\begin{tabular}{cccccccc} 
		\hline
		Fried Parameter & Compressive Rate & Blurred Image &  Restored Image & TOR & NOR & TNR & NNR \\
		\hline
		& 0.1 &  & & & & 0.2910±0.0075 & 0.3037±0.0105\\
             & 0.3 &  & &   &  & 0.2868±0.0066 & 0.3056±0.0091\\
             6 cm& 0.5 & 0.1952±0.0021 & 0.3094±0.0039 & 0.3076±0.0069 & 0.3019±0.0114 & 0.2841±0.0069 & 0.3055±0.0081\\
             & 0.7 &  & &   &  & 0.2809±0.0078 & 0.3053±0.0080\\
             & 0.9 &  & &   &  & 0.2756±0.0073 & 0.3073±0.0091\\
		\hline
		& 0.1 &  & & & & 0.4352±0.0072 & 0.4531±0.0095\\
             & 0.3 &  & &   &  & 0.4258±0.0074 & 0.4531±0.0097\\
             12 cm& 0.5 & 0.2493±0.0035 & 0.4360±0.0070 & 0.4610±0.0074 & 0.4515±0.0100 &0.4191±0.0077 & 0.4494±0.0111\\
             & 0.7 &  & &   &  & 0.4138±0.0075 & 0.4479±0.0114\\
             & 0.9 &  & &   &  & 0.4050±0.0096 & 0.4509±0.0111\\
		\hline
		& 0.1 &  & & & &0.5090±0.0139 & 0.5248±0.0160\\
             & 0.3 &  & &   &  & 0.5032±0.0138 & 0.5211±0.0175\\
             18 cm& 0.5 & 0.2750±0.0071 &0.5065±0.0155 & 0.5368±0.0153 & 0.5242±0.0160 &0.4971±0.0123 & 0.5162±0.0191\\
             & 0.7 &  & &   &  & 0.4863±0.0115 & 0.5130±0.0202\\
             & 0.9 &  & &   &  & 0.4731±0.0119 &0.5153±0.0204\\
		\hline
	\end{tabular}
\end{table*}

\begin{figure*}
\centering
\subfigure[The original image.]{
\begin{minipage}[t]{1\columnwidth}
\centering
\includegraphics[width=1\columnwidth]{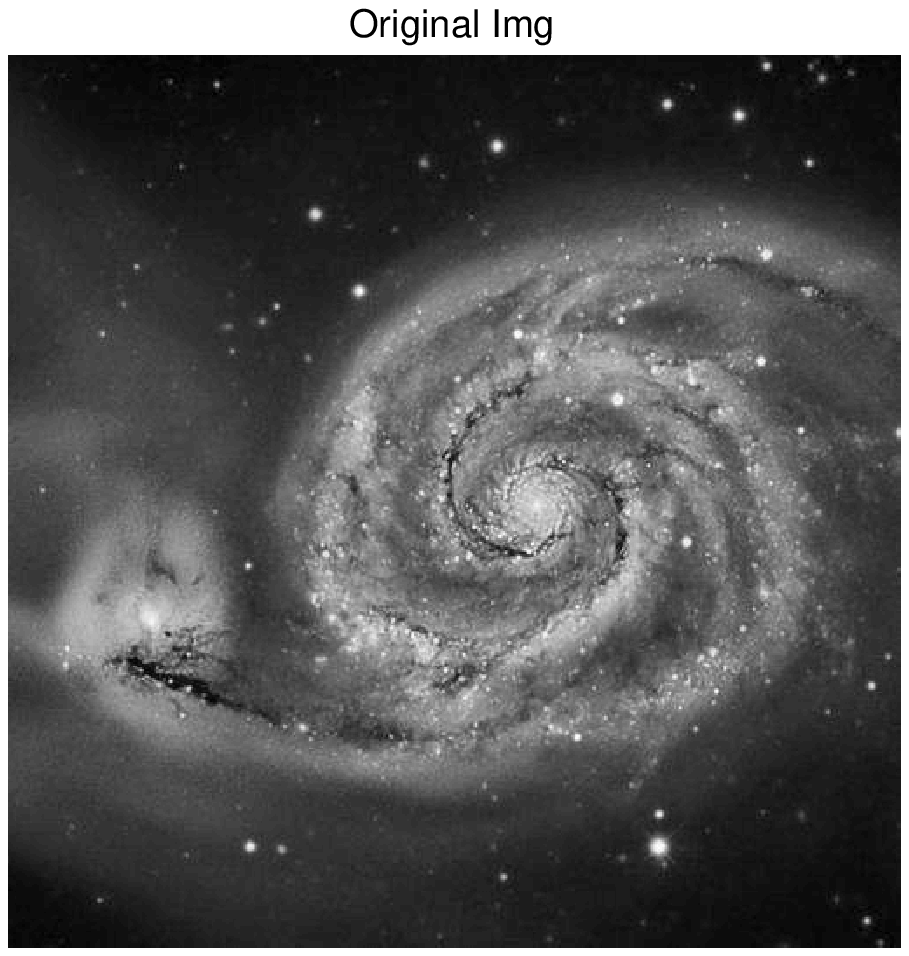}
\end{minipage}%
}%
\subfigure[Image blurred by atmospheric turbulence with Fried parameter of 6 cm.]{
\begin{minipage}[t]{1\columnwidth}
\centering
\includegraphics[width=1\columnwidth]{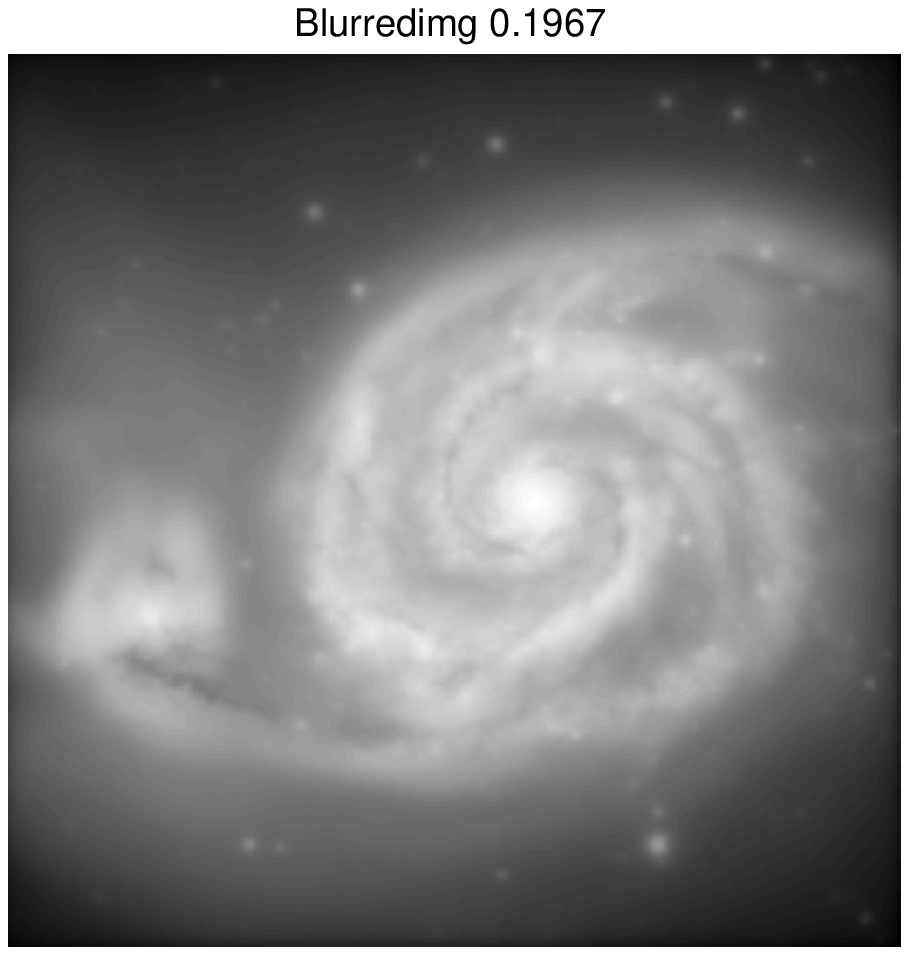}
\end{minipage}%
}\\
\subfigure[Images restored from original wavefront with the classsical method.]{
\begin{minipage}[t]{1\columnwidth}
\centering
\includegraphics[width=1\columnwidth]{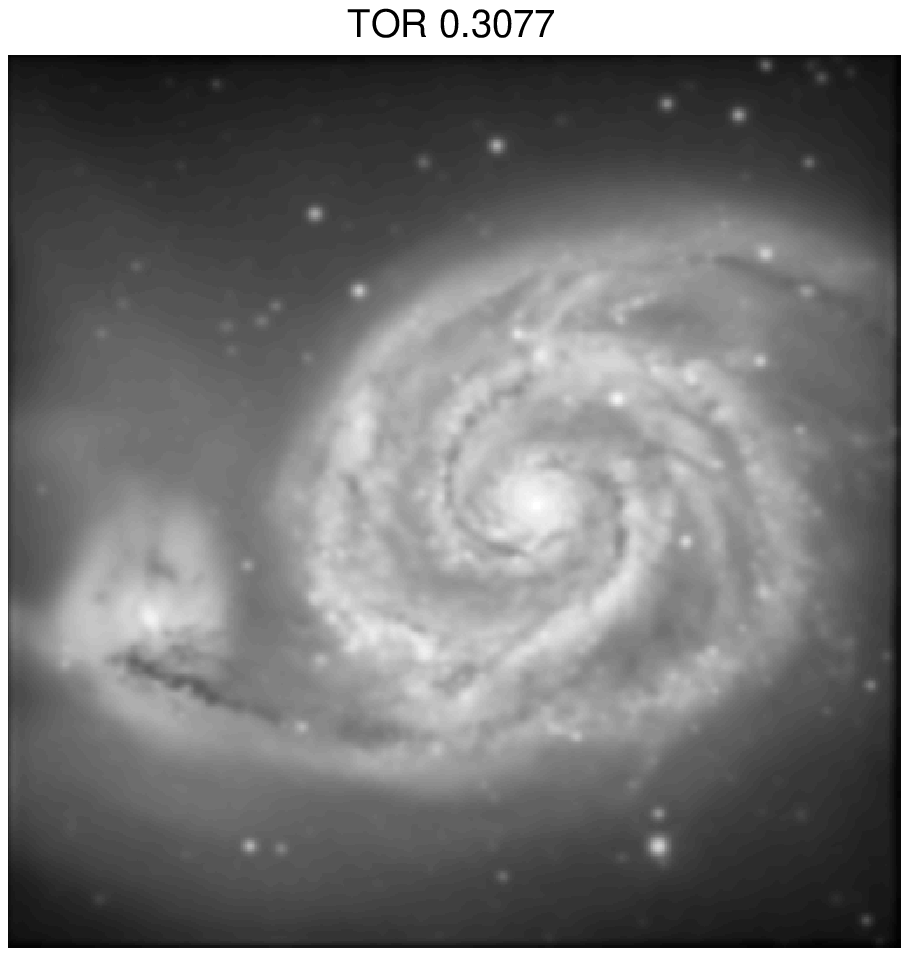}
\end{minipage}
}%
\subfigure[Images restored from original wavefront with the neural network.]{
\begin{minipage}[t]{1\columnwidth}
\centering
\includegraphics[width=1\columnwidth]{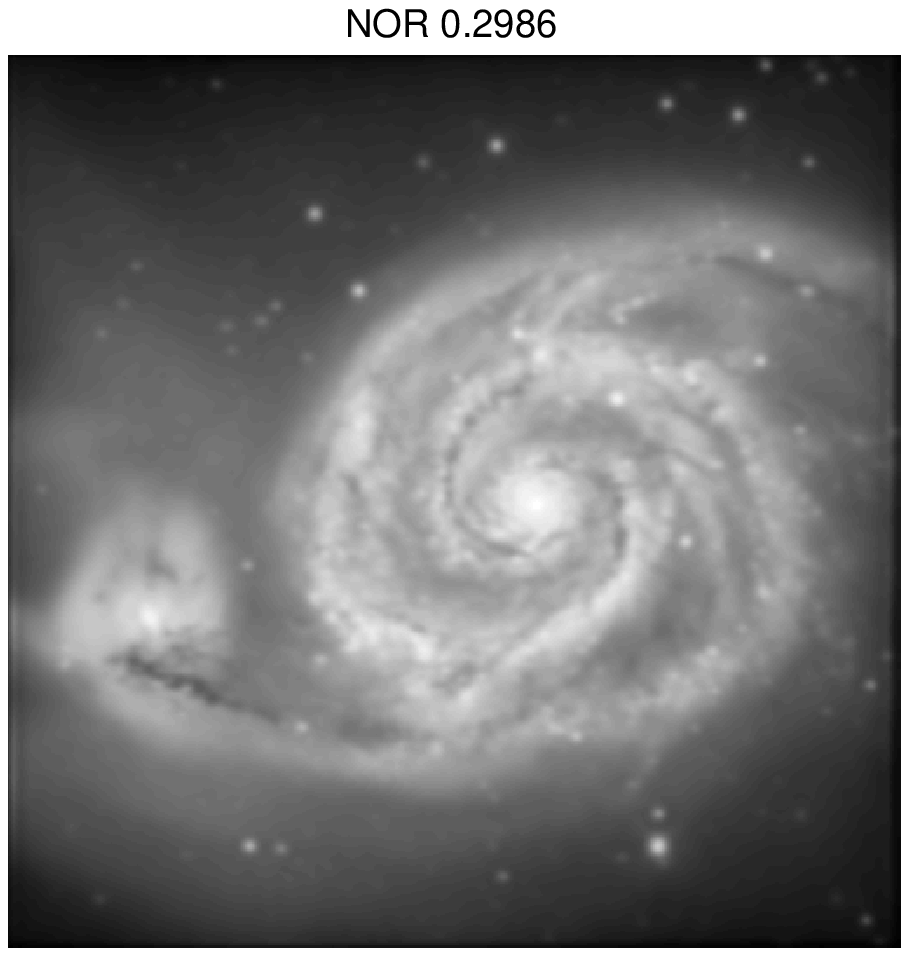}
\end{minipage}
}\\
\subfigure[Images restored from noisy wavefront with the classsical method.]{
\begin{minipage}[t]{1\columnwidth}
\centering
\includegraphics[width=1\columnwidth]{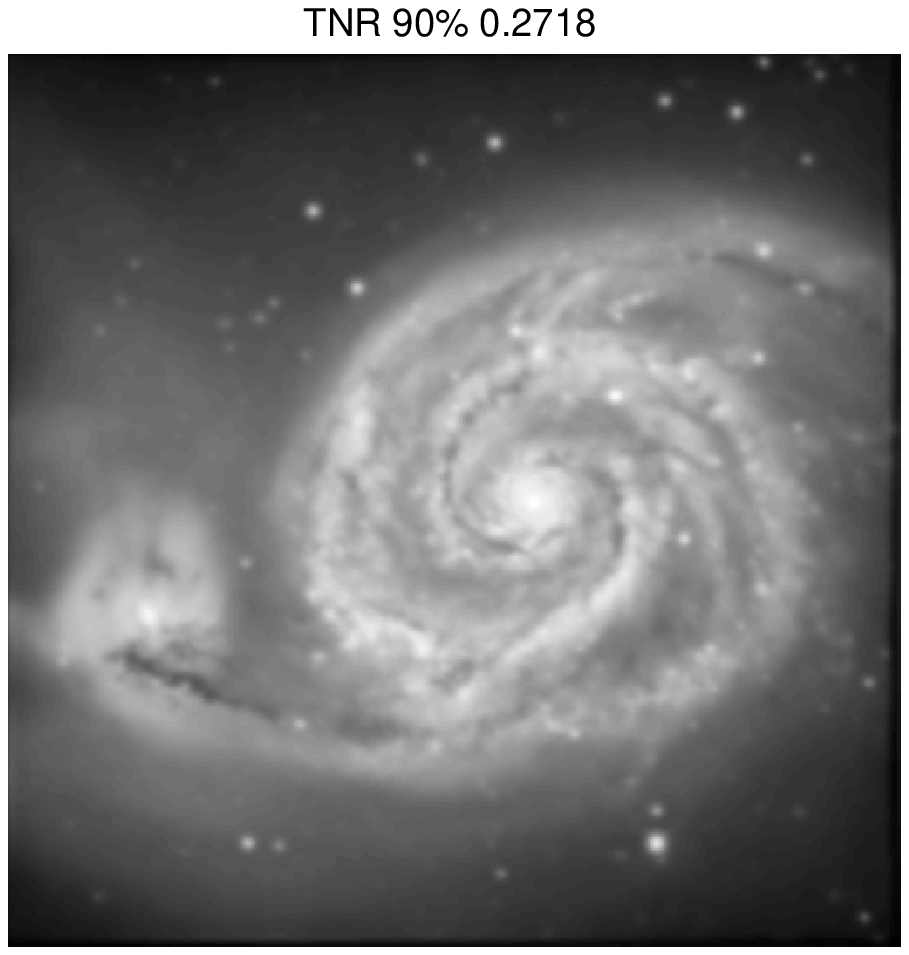}
\end{minipage}
}%
\subfigure[Images restored from noisy wavefront with the neural network.]{
\begin{minipage}[t]{1\columnwidth}
\centering
\includegraphics[width=1\columnwidth]{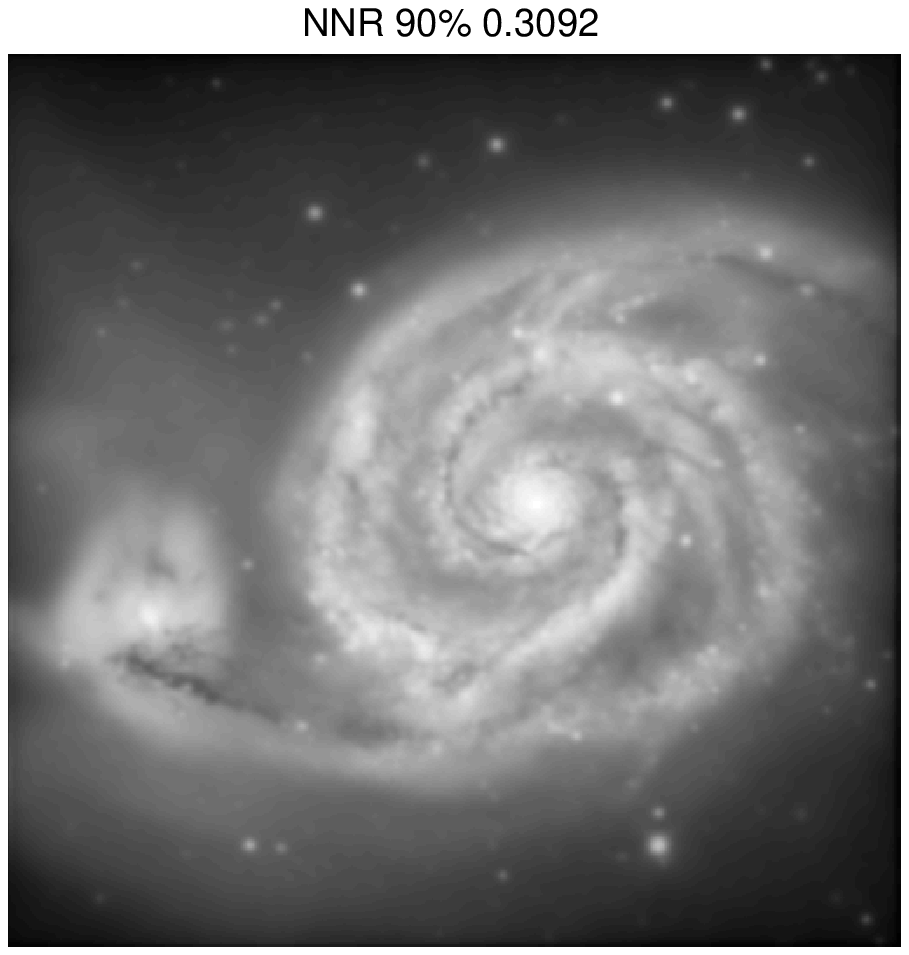}
\end{minipage}
}%
\centering
\caption{(a) is the original image. (b) is the image blurred by random atmospheric turbulence with $r_0$ of 6 cm. (c) is the restored image with the PSF estimated from wavefronts obtained from classical method.  (d) is the restored image with the PSF estimated from wavefronts obtained from the neural network. (e) is the restored image with the PSF estimated from noisy wavefronts from classical method. (f) is the restored image with the PSF estimated from noisy wavefronts from the neural network.}
\label{fig:8}
\end{figure*}

Blurred images and restored images are shown in figure \ref{fig:8}. img stands for the original image. Blurred image is one frame of images blurred by the atmospheric turbulence phase screen. TOR stands for restored images with PSFs estimated from original wavefront with the classical method. NOR stands for restored images with PSFs estimated from original images with the neural network. TNR stands for restored images from noisy wavefront with the classical method and NNR stands for restored images from noisy wavefront with the neural network. We can find that even with small amount of slope measurements, our CS-WFS can still obtain wavefront that can effectively restore images. Because the uncertainty brought by the slope measurements selection step would reduce the performance of the DWFS, we also estimate the SSIM of NGC 4192 for different slope selection situations as defined in Section \ref{sec:CSWFResult}. The results are shown in table \ref{tab:SSIM2}. The results are similar to wavefront reconstruction results. A trained CS-WFS with DWFS can effectively restore images with sparsely sampled wavefronts.\\

\begin{table*}
	\centering
	\caption{SSIM of restored images with different percentages of selected measurements for wavefront measurements with different noise level. Fried parameter stands for Fried parameter in centimetre. Selection rate stands for percentage of removed slope measurements (comparing with noisy slope measurements) that are used for wavefront reconstruction. Noise level with different number stands for wavefront measurements with different percentage of noisy slope measurements.}
	\label{tab:SSIM2}
	\begin{tabular}{cccccccc} 
		\hline
		Fried Parameter & Selection Rate & Noise level 0 & Noise level 0.1 & Noise level 0.3 & Noise level 0.5 & Noise level 0.7 & Noise level 0.9 \\
		\hline
		& $60\%$ &  &0.3030±0.0108 & 0.3048±0.0101 & 0.3054±0.0086 & 0.3058±0.0081 & 0.3054±0.0082\\
             & $80\%$ &  & 0.3034±0.0107& 0.3053±0.0096  & 0.3056±0.0084  & 0.3051±0.0078 & 0.3059±0.0082\\
            6 cm& $100\%$ & 0.3019±0.0114 & 0.3037±0.0105 & 0.3056±0.0091  & 0.3055±0.0081 & 0.3053±0.0080 & 0.3073±0.0091\\
              & $120\%$ &&  0.3040±0.0104 & 0.3058±0.0085 & 0.3050±0.0085 &0.3063±0.0085 & \\
		\hline
		& $60\%$ &  & 0.4525±0.0096 & 0.4535±0.0096 & 0.4530±0.0097 & 0.4509±0.0103 & 0.4493±0.0105\\
             & $80\%$ &  & 0.4528±0.0096 & 0.4532±0.0098 & 0.4510±0.0108 & 0.4488±0.0109 & 0.4479±0.0109\\
             12 cm& $100\%$ & 0.4515±0.0100 & 0.4531±0.0095 & 0.4531±0.0097 & 0.4494±0.0111 & 0.4479±0.0114 & 0.4509±0.0111\\
             &$120\%$ &  & 0.4533±0.0096 & 0.4521±0.0099 & 0.4485±0.0111 & 0.4489±0.0120 & \\
		\hline
		& $60\%$ &  &0.5249±0.0158 & 0.5237±0.0167 & 0.5221±0.0172 & 0.5177±0.0185 & 0.5140±0.0202\\
             & $80\%$ &  & 0.5248±0.0160 & 0.5227±0.0172  & 0.5194±0.0182 & 0.5138±0.0191 & 0.5134±0.0203\\
            18 cm& $100\%$ & 0.5242±0.0160 & 0.5248±0.0160 &  0.5211±0.0175 & 0.5162±0.0191 & 0.5130±0.0202 & 0.5153±0.0204\\
             &$120\%$ &  & 0.5248±0.0161 & 0.5194±0.0184  & 0.5141±0.0197 & 0.5131±0.0197 & \\
		\hline
	\end{tabular}
\end{table*}

\section{Conclusions and future work}
\label{sec:con} 
In this paper, we propose a compressive Shack-Hartmann wavefront sensing method (CS-WFS). Our method uses a drop-out layer to simulate the compressive sensing process during training. After training, the CS-WFS can reconstruct wavefront with only several to tens of percentage of sub-apertures. We integrate the CS-WFS with the DWFS to test the performance of our method. The results show that our method is robust and can fast estimate wavefronts effectively.\\
The CS-WFS provide an effective way to measure high order wavefronts, which would further increase performance of high-order adaptive optics systems \citep{fusco2006high, roberts2012results, shumko2014308, rao2015second} or PSF reconstruction algorithms \citep{martin2016psf, beltramo2019prime, fetick2019physics, fusco2020reconstruction}. Besides, the compressive Shack-Hartmann wavefront sensing algorithm can reduce the requirement of brightness of guide stars in an adaptive optics system. Wide field adaptive optics system, such as ground layer adaptive optics system could then use natural guide stars as wavefront references for observations \citep{rigaut2002ground, nicolle2004ground, travouillon2004ground, chun2016imaka, carbillet2017anisoplanatic, jia2018ground, lu2018ground}. In our future works, we will integrate the CS-WFS with the deep learning based PSF model \citep{jia2020psf} to develop high-order PSF reconstruction method and we will also further investigate applications of the compressive wavefront sensing in wide field adaptive optics systems.

\section*{Acknowledgements}
Peng Jia would like to thank Professor Dingqiang Su from Nanjing University for guidance of the principle of Shack-Hartmann wavefront sensor. Peng Jia would like to thank Professor Xiangqun Cui from Nanjing Institute of Astronomical Optics \& Technology, China Academy of Sciences for guidance on Shack-Hartmann wavefront sensor in ground layer adaptive optics systems for large telescopes. Peng Jia would also like to thank Professor Yong Zhang from Nanjing Institute of Astronomical Optics \& Technology, China Academy of Sciences for introducing of the Shack-Hartmann wavefront sensor in the Large Sky Area Multi-Object Fibre Spectroscopic Telescope. All authors are grateful for kindly suggestions provided by the reviewer. Data resources are supported by China National Astronomical Data Center (NADC) and Chinese Virtual Observatory (China-VO). This work is supported by Astronomical Big Data Joint Research Center, co-founded by National Astronomical Observatories, Chinese Academy of Sciences and Alibaba Cloud.\\

This work is supported by National Natural Science Foundation of China (NSFC) (11503018, 61805173), the Joint Research Fund in Astronomy (U1631133) under cooperative agreement between the NSFC and Chinese Academy of Sciences (CAS). Authors acknowledge the French National Research Agency (ANR) to support this work through the ANR APPLY (grant ANR-19-CE31-0011) coordinated by B. Neichel. This work is also supported by Shanxi Province Science Foundation for Youths (201901D211081), Research and Development Program of Shanxi (201903D121161), Research Project Supported by Shanxi Scholarship Council of China (HGKY2019039), the Scientific and Technological Innovation Programs of Higher Education Institutions in Shanxi (2019L0225). \\ 

\section*{Data Availability}

The code used in this paper is written in Python programming language (Python Software Foundation) with the package Pytorch and the complete version of our code can be downloaded from PaperData Repository Powered by China-VO (\href{CS-WFS}{https://nadc.china-vo.org/article/20200722160959?id=101045}) with a DOI number of 10.12149/101045.\\



\bibliographystyle{mnras}
\bibliography{CS-WFS} 



%




\bsp	
\label{lastpage}
\end{document}